%Paper: hep-th/9212050
%From: bazhanov@pell.anu.edu.au (Vladimir Bazhanov)
%Date: Wed, 9 Dec 92 02:05:26 EST
%Date (revised): Wed, 9 Dec 92 22:17:04 EST

%Plain TeX, 3 figures with instructions included after `\end' command
\magnification=1200
\baselineskip 12pt
\def\w{\omega}
\def\s{\sigma}
\def\a{\alpha}
\def\b{\beta}
\def\g{\gamma}
\def\d{\delta}

\def\l{\lambda}
\def\m{\mu}

\def\os{\overline{s}}
\def\t{\theta}

\def\D#1{\Delta(#1)}

\def\BW{Boltzmann weight}

\def\IRC{interaction-round-a-cube}
\def\PFS{partition function per site}
\def
\ZM{Zamolodchikov model}
\def\trace{\mathop{\rm Tr}\nolimits}

\def\L{{\cal L}}
\def\ss{s_0,\ldots ,{\overline s}_0}

\def\fig #1#2{}
\def\f{\phi}

{\nopagenumbers
\line{\hfill 25 July 1992}
\line{\hfill SMS-079-92/MRR-020-92}
\line{\hfill hep-th/9212050}
\vskip 1cm
\centerline{\bf Star-Triangle Relation for a Three Dimensional Model.
}
\vskip 1 cm
\centerline{V.\ V.\ Bazhanov\footnote{${}^\sharp$}{On leave
of absence from the Institute for High Energy Physics,
Protvino, Moscow Region, 142284, Russia.}}
\bigskip\sl
\centerline{Mathematics \& Theoretical Physics, IAS,}
%\centerline{Institute for Advanced Study,}
\centerline{Australian National University,}
\centerline{GPO Box 4, Canberra, ACT 2601, Australia}
%\centerline{and}
%\centerline{Institute for High Energy Physics,}
%\centerline{Protvino, Moscow Region, 142284, Russia.}\rm
\rm\bigskip
\centerline{and}
\bigskip
\centerline{R.\ J.\ Baxter}\sl\bigskip
\sl\centerline{Mathematics \& Theoretical Physics, IAS,}
%\centerline{Institute for Advanced Study,}
\centerline{Australian National University,}
\centerline{GPO Box 4, Canberra, ACT 2601, Australia}
%\bigskip
\centerline{and}
%\bigskip
\centerline{Isaac Newton Institute for Mathematical Sciences,}
\centerline{Cambridge University, England}
\rm

\vskip 1 cm

{\bf Abstract.} The solvable $sl(n)$-chiral Potts model can
be interpreted as a three-dimen\-sional
lattice  model with local interactions.
To within a minor modification of the boundary conditions
it is an Ising type model on the body centered cubic lattice
with two- and three-spin
interactions. The corresponding local
Boltzmann weights obey a number of simple relations, including a
restricted star-triangle relation, which is a modified version of the
well-known  star-triangle relation appearing in two-dimen\-sional models.
We show that these relations lead to remarkable symmetry properties
of the Boltzmann weight function of an elementary cube of the lattice,
related to spatial symmetry group of the cubic lattice.
These symmetry properties allow one  to prove the commutativity of the
row-to-row transfer matrices, bypassing   the tetrahedron relation.
The partition function per site for the infinite lattice is calculated
exactly.


\vfill\eject}

\beginsection{ Introduction and Summary}

\bigskip

There is a large numbers of solvable models in statistical mechanics
where the bulk free energy and possibly other
quantities such as order parameters and the correlation length can be
calculated exactly.
Mostly these models  are two-dimensional, and only a few
three-dimen\-sional examples are known. The first such
model was the two-dimen\-sional Ising model, solved by Onsager in 1944 [1].
Then, it took some two decades  to realize that two important ingredients
of the Ising model --- the star-triangle relation (STR) and the resulting
commutativity
of the transfer matrices --- can be used to solve the other
two-dimen\-sional models [2,3].

There is now quite rich theory of solvable two-dimensional lattice
models.
There are several different (but related) methods, mostly based on the
commutativity of the transfer matrices, and thereby on the Yang-Baxter
relation which have been developed
(for a review see [4,5]).

Can we find three-dimen\-sional models with commuting transfer
matrices?
It is known that
the tetrahedron relation [6] replaces the Yang-Baxter relation as a
commutativity condition [7,8] for the three-dimensional cubic lattice.
This relation contains thousands of distinct algebraic equations, and,
obviously, it is very  difficult to solve them. The only non-trivial solution
obtained by direct analysis of these equations known so far is that of
Zamolodchikov [6,9].
In fact, even in the simplest two state spin case one has to solve
$2^{14}$ equations, instead of $2^6$ in two dimensions. The symmetry
properties could slightly reduce this number, but  the jump in
complexity is still enormous. Therefore one would like to find an
alternative approach to the commutativity which would be based on more
simple algebraic relations.

In  the present paper we discuss one such scheme. We consider the solvable
\IRC\ model [10] on the cubic lattice with $N$-valued spins ($N\ge2$) at
each site. To within a minor modification of the boundary condition it
is equivalent to the $sl(n)$-chiral Potts model [11]  (this model has
an interesting history which can be traced in [12-17]). On the other
hand it can be regarded as a multistate generalization
of the Zamolodchikov model [6], reducing to it when $N=2$.
The Boltzmann weight function  of  eight
corner spins around a cube have a very special
form, such that by introducing an auxiliary center spin for each cube
the model can be viewed as an Ising-type
one on the body centered cubic  lattice
with two- and three-spin interactions only. The corresponding
Boltzmann weights are of course not arbitrary,
and   obey a number of  simple relations. The most
complicated among them is a ``restricted star-triangle relation'' (RSTR)
which relates two- and three-spin weights. Remarkably enough this
relation is a
simplified version of the two-dimensional star-triangle relation (for
the two-state spin case it can be obtained simply
by specialization of rapidities
in the STR of the 2d Ising model).

Starting from these relations we derive symmetry properties
of a Boltzmann weight function  of the elementary cube of the lattice
under  transformations from the symmetry group of the cube.
As in the Zamolodchikov model this Boltzmann weight function depends
on three parameters which can conveniently be chosen to be the
dihedral angles $\t_1,\t_2,\t_3$
between three ``rapidity planes'' passing through the
cube. The symmetry properties are entirely consistent with such
geometric interpretation of these angles.

The
implication of the symmetry properties is twofold. First, they contain
the three-dimensional star-star relation
[10,18]\footnote{${}^1$}{Interestingly, this relation corresponds to the
``central inversion'' of the cubic lattice.
}
 which results in the
commutativity of the transfer matrices. The other symmetry relations imply
certain symmetry properties of the partition function per site
for the
lattice infinite in the all three directions. These
two features --- commutativity and symmetry --- allow us to find the
partition function per site
extending  the $N=2$ Zamolodchikov case calculations of [18] to the
general $N$ case. The result is quite surprising:
$$
\log \kappa_N(\t_1,\t_2,\t_3)={2(N-1)\over N}\log\kappa_2(\t_1,\t_2,\t_3)
$$
where $\kappa_N$ is the partition function per site for the $N$-state model.
In addition we show that there exists a three-dimensional
free boson (Gaussian)
model whose partition function per site is given by
$$
\log \kappa_B(\t_1,\t_2,\t_3)=-2\log\kappa_2(\t_1,\t_2,\t_3)
$$
This means that $\kappa_N$ can be expressed as a (rational) power of
some free boson determinant. It would be very interesting to understand
the reason for this fact.

The sequence of our working  is illustrated on Fig.1. We hope this
will assist the reader.

\def\odd{\Phi}
 \def\equalign{\eqalign}
\def\equalignno{\eqalignno}  
\def\nt{\noindent}
\def\h{{1\over 2}}

\beginsection{\bf 1. Formulation of the model\hfill}

\nt 1.1\quad The interaction-round-a-cube model

\bigskip
Consider a simple cubic lattice $\cal L$ of $M$ sites with the periodical
boundary conditions in each direction.  At each site of $\cal L$ place a
spin
variable $s$ taking $N \geq 2$ distinct values $s = 0,\ldots , N
-
1$, and allow all possible interactions of the spins within each elementary
cube.  The partition function reads
$$Z = \sum_{\rm{spins}} \ \prod_{\rm{cubes}}\,V_1(s_0 \vert {\overline
s}_1,\,{\overline s}_2,\,{\overline s}_3\vert\,s_1,\, s_2,\,s_3\vert
{\overline
s}_0) \eqno(1.1)$$
where $s_0,\ldots ,{\overline s}_0$ are the eight spins of the cube
arranged as
in Fig.~2, and $V(s_0\,\vert {\overline s}_1,\,{\overline
s}_2,\,\break{\overline s}_3\vert s_1,\,s_2,\,s_3\vert \,{\overline s}_0)$
is
the Boltzmann weight of the spin configuration $s_0,\ldots ,{\overline
s}_0$.
Note that the spins $s_0,\,s_1,\,s_2,\,s_3,\,{\overline s}_1,\,{\overline
s}_2,\,{\overline s}_3 ,\,{\overline s}_0$ correspond respectively to $a$,
$b$,
$c$, $d$, $e$, $f$, $g$, $h$ in Fig.~2 of reference [10].  The product is
over
all elementary cubes in $\cal L$.

Taking the lattice to have $m$ horizontal layers and letting $\phi_i$
denote
all spins in layer $i$, one can rewrite (1.1) as
$$Z = \sum_{\phi_1}\ \sum_{\phi_2} \cdots
\sum_{\phi_m}\,T_{\phi_1\phi_2}T_{\phi_2\phi_3} \cdots T_{\phi_m\phi_1} =
{\rm{Tr}}\,T^m\ \ \ ,\eqno(1.2)$$
where $T$ is a layer-to-layer transfer matrix whose elements are the
products
of all the $V$ functions of cubes between two adjacent layers.  Clearly,
$T$
depends on the Boltzmann weight function $V$, so we can write it as $T(V)$.

Obviously, we cannot solve the model (1.1) for an arbitrary Boltzmann
weight
function $V$.  What we apparently can do is to solve a restricted class of
models whose transfer matrices form commuting families, i.e., such that any
two
transfer matrices $T(V)$ and $T(V^{\prime})$ belonging to the same family
commute,
$$\left[ T(V),\,T(V^{\prime})\right] = 0\ \ \ .\eqno(1.3)$$

In all known cases this commutativity condition can always be reformulated
as a
relation for the local Boltzmann weight functions $V$ and $V^{\prime}$.  In
two
dimensions the corresponding relation is well known:  it is the Yang-Baxter
equation for the local Boltzmann weights.  This equation involves three
different Boltzmann weight functions.  A straightforward
generalization [7,8] of
this construction for a 3-dimensional lattice leads to the tetrahedron
equation
[6], involving four different Boltzmann weight functions.  Even in the
simplest two-valued spin case this equation involves thousands of distinct
relations and, obviously, is very difficult to solve.  The only known
non-trivial solution
obtained from the direct analysis of these relations is the Zamolodchikov's
one
[6,9].

In this paper we present a somewhat different approach to the commutativity
in three dimensions, which could, however, be applied to the Zamolodchikov
model as well. We consider the \IRC\    model of ref [10]
where the Boltzmann weight function  of  eight
corner spins around a cube, $V$,  has a very special
form, such that by introducing an auxiliary center spin for each cube
the model can be viewed as an Ising-type
one on the body centered cubic  lattice
with two- and three-spin interactions only.
The corresponding 2-spin
and
3-spin Boltzmann weights obey a number of simple relations including ``a
restricted star-triangle relation'' (RSTR) which allow us to prove the
commutativity of
the transfer matrices and calculate the partition function per site by use
of the
symmetry properties of the model.

Before ending this subsection note that there are simple transformations of
$V$
that do not change (1.1).  For instance, if we multiply $V(s_0,\ldots
,{\overline s})$ by $F(s_0,\,{\overline s}_2,\,s_1,\,{\overline s}_3)
/\break F({\overline s}_1,\,s_3,\,{\overline s}_0,\,s_2)$ then each
horizontal
face of $\cal L$ acquires an $F$-factor from the cube below it, and a
canceling
$(1/F)$ factor from the cube above.  The effect on the transfer matrix $T$
is to
apply a diagonal similarity transformation.  Provided $F$ is the same for
$T(V)$
and $T(V^{\prime})$ the commutation relation is unaffected.  This is an
example
of a ``face-factor'' transformation.  Similarly, one could apply
``edge-factor''
and ``site-factor'' transformations that leave (1.1) and (1.3) unchanged.

\bigskip\bigskip
\nt 1.2\quad The Boltzmann weights

\bigskip
Let $a$, $b$, $c$ be integers and $v$ be a complex variable.  Introduce the
following notation
$$\equalignno{\omega = \exp (2\pi i/N)\ \ ,\ \ &\omega^{1/2} = \exp
(i\pi/N)\ \
,&(1.4)\cr
\noalign{\vskip8pt}
\g(a,\,b) = \omega^{ab}\ \ ,\ \ \ &\odd(a) = (\omega^{1/2})^{a(N+a)}\ \
,&(1.5)\cr
\noalign{\vskip8pt}
{}\Delta(v) = (1 - v^N)^{1/N}&\ \ \ .&(1.6)\cr}$$

Note that
$$\equalignno{\g(a,\,b) = \g(b,\,a) &= \g(a + N,\,b)\ \ \ ,\cr
\noalign{\vskip8pt}
\odd (a + N) &= \odd (a)\ \ \ .&(1.7)\cr}$$

Define a function $w(v,\,a)$ such that
$${{w(v,\,a)}\over{w(v,\,0)}} = \left(\Delta(v)\right)^a\ \prod^a_{j=1} (1
-
\omega^j\,v)^{-1}\ \ \ .\eqno(1.8)$$
Obviously, the LHS of this equation is a multivalued function of $v$, but
it is
a singlevalued function of a point $\left(v,\,\Delta(v)\right)$ on the
algebraic
curve $$v^N + \Delta^N = 1\ \ \ .\eqno(1.9)$$
Below we shall interpret $w$ in this latter way, suppressing, however, an
explicit dependence on the phase of $\Delta(v)$ in the arguments of $w$.

Now, fix four complex parameters $p$, $p^{\prime}$, $q$, $q^{\prime}$, and
define
$$v_1 = q^{\prime}/(\omega\,p^{\prime})\ \ ,\ \ v_2 = q^{\prime}/p\ \ ,\ \
v_3
= p/q\ \ ,\ \ v_4 = p^{\prime}/q\ \ .\eqno(1.10)$$
Obviously
$$\omega\,v_1\,v_4 = v_2\,v_3\ \ \ .\eqno(1.11)$$

The Boltzmann weight function of the model is given by the equations (2.9),
(2.10) of ref.[10].  Taking into account equation (1.18) (given below)
and
omitting equivalence transformation factors (which do not change the
partition
function) one can write it in the form
$$\equalignno{{}&V(s_0\vert {\overline s}_1,\,{\overline s}_2,\,{\overline
s}_3
\vert s_1,\,s_2,\,s_3 \vert {\overline s}_0 \Vert v_1,\,v_2,\,v_3,\,v_4) =
\cr
\noalign{\vskip8pt}
{}&= \rho\,\sum^{N-1}_{\sigma = 0}\,{{w(v_2,\,s_1 - {\overline s}_2 +
\sigma)\,w\,(v_3,\,s_3 - {\overline s}_0 -
\sigma)\,\g(s_0,\sigma)\g(\os_0,\s)}
\over{w (v_1,\,{\overline s}_3 - s_0 +
\sigma)\,w\,(v_4,\,{\overline s}_1 - s_2 - \sigma)\,\g(s_2,\sigma)
\g(\os_2,\s)}}&(1.12)\cr}$$
where $\rho$ is a normalization factor depending on $v_1,\ldots ,v_4$.
Note that the Boltzmann weight function (1.12) describes a very special
type of
interaction of eight spins around the cube.  A typical cube with its centre
spin, $\sigma$, is shown in Figure 2.  There are three-spin interactions on
the
shaded triangles, described by $w(v,\,a)$ or by $1/w(v,\,a)$ in (1.12).
There
are also two-spin interactions such as $\g(s_0,\,\sigma)$, or
$1/\g(s_2,\,\sigma)$
associated with the edges linking $\sigma$ to $s_0$, ${\overline s}_0$,
$s_2$,
${\overline s}_2$ (these edges are denoted by heavy lines in Figure 2.).
In
addition, the three-spin and two-spin interaction weights $w$ and $s$ obey
a
number of important relations which we now discuss.  These relations,
rather
than the explicit form of $w$ and $\g$ given by (1.5) and (1.8), enable us
to
execute the program stated in the introduction and calculate the partition
function
per
site
in the thermodynamic limit.

\bigskip\bigskip
\nt 1.3\quad Restricted star-triangle relation.

\bigskip
Consider the following automorphism of the curve (1.9)
$$\equalignno{(v,\,\Delta(v)) &\rightarrow ({\tilde v},\,\Delta ({\tilde
v}))\ \
\ ,&(1.13)\cr
\noalign{\vskip8pt}
{\tilde v} = {1\over{\omega v}}\ \ ,\ \ \Delta ({\tilde v}) &=
{{w^{-1/2}\Delta
(v)}\over v}.&(1.14)\cr}$$

One can easily show that $\g(a,\,b)$, $\odd (a)$ and $w(v,\,a)$ satisfy the
following properties
$$\equalignno{\g(a,\,b + c) &= \g(a,\,b)\ \g(a,\,c)&(1.15)\cr
\noalign{\vskip8pt}
\sum^{N-1}_{b=0}\ \g(a,\,b)\ \g(-b,\,c) &= N\,\delta_{ac}&(1.16)\cr
\noalign{\vskip8pt}
\odd (a + b) &= \odd (a)\ \odd (b)\ \g(a,\,b)&(1.17)\cr
\noalign{\vskip8pt}
{{w(v,\,a)\,w({\tilde v}, - a)}\over{w (v,\,0)\ w({\tilde v},\,0)}} &=
\odd^{-1}(a)&(1.18)\cr}$$
where $\tilde v$ is related to $v$ by equation (1.14).
Moreover
these functions satisfy two less trivial relations.  For the moment let
$v_1$,
$v_2$, $v_3$, $v_4$ denote four arbitrary complex variables, then
$$F_1(v_1,\,v_2 \vert a,\,b) \equiv \sum^{N-1}_{\ell = 0}\
{{w(v_2,\,a - \ell)}\over{w(v_1, -\ell)\,\g(b,\,\ell)}} =
\varphi_1(v_1,\,v_2)\
{{w(v^{\prime}_2, - b)\ w(v_2/(\omega v_1),\,a)}\over{w(v^{\prime}_1,\,a -
b)}}\
\ \ ,\eqno(1.19)$$

$$F_2(v_3,\,v_4 \vert a,\,b) \equiv \sum^{N-1}_{\ell =
0}\,{{w(v_3,\,-\ell)\,\g(b,\,\ell)}\over{w(v_4,\,a - \ell)}} =
\varphi_2(v_3,\,v_4)\ {{w(v^{\prime}_3,\,a -
b)}\over{w(v^{\prime}_4,\,-b)\,w(v_4/v_3,\,a)}}\ \ ,\eqno(1.20)$$
where $\varphi_1$ and $\varphi_2$ are scalar functions and

$$\equalignno{
v^{\prime}_1 = {v_2\,\Delta\,(v_1)\over\w v_1\,\Delta\,(v_2)}\ \ ,\qquad
&\Delta(v^{\prime}_1) = {\Delta\,(v_2/(\w v_1))\over\Delta\,(v_2)}\ \ \
,&(1.21a)\cr
\noalign{\vskip8pt}
v^{\prime}_2 = {{\Delta(v_1)}\over{\Delta(v_2)}}\ \ \qquad ,
\qquad&\Delta(v^{\prime}_1) = {{v_2\Delta(v_2/(\w v_1))}\over{\Delta(v_2)}}\
\ \
,&(1.21b)\cr
\noalign{\vskip8pt}
v^{\prime}_3 = {{v_4\Delta(v_3)}\over{v_3\Delta(v_4)}}\ \ \quad,\qquad
&\Delta(v^{\prime}_3) = {{\Delta(v_4/v_3)}\over{\Delta(v_4)}}\ \ \
,&(1.21c)\cr
\noalign{\vskip8pt}
v^{\prime}_4 = {{\Delta(v_3)}\over{\omega\Delta(v_4)}}\ \ \quad
,\qquad&\Delta(v^{\prime}_4) = {{v_3\Delta(v_4/v_3)}\over{\Delta(v_4)}}\
\ \
.&(1.21d)\cr}$$
Each of the relations (1.19), (1.20) is a corollary of another and the
properties (1.15) - (1.18).  The phases of $\Delta(v_4/v_3)$ and
$\Delta(v_2/(\omega\,v_1))$ can be chosen arbitrarily since they cancel out
the
RHSs of (1.19) and (1.20).

The reader may have noticed that relation (1.19) (or (1.20)) is a
particular case of the usual star-triangle relation.   Indeed, the LHS
of (1.19) is the sum of the product of the three functions (each
depending on two spins) while the RHS is  the product of three such function.
Unlike the usual star-triangle relation there is some asymmetry
in the LHS of (1.19): the function $\g$ does not depend on any
continuous parameters. It is quite possible that (1.19) is a
particular case of a more general relation and $\g$ is just a limiting
value of a more complex function. In fact, this is exactly so for
$N=2$ when (1.19) and (1.20) can be obtained be a  specialization of
rapidities in the the star-triangle relation of the two-dimensional
Ising model [1]. This is the reason why we call the (1.19), (1.20) as the
``restricted star-triangle relations''.

\beginsection{2.  The symmetry relations}

\bigskip
\nt{2.1\quad The cube symmetry}

\bigskip
Consider the cube ${\cal C}$ shown in Fig. 2.  The eight spins
$s_0,\ldots
,{\overline s}_0$ at the corners of ${\cal C}$ can be grouped into four
ordered
pairs $$d_j = (s_j,\,{\overline s}_j)\ \ \ ,\ \ \ j = 0,\ldots ,3
\eqno(2.1)$$
corresponding to four diagonals $d_0,\ldots ,d_j$, which we assume to be
oriented and directed from $s_j$ to ${\overline s}_j$, $j = 0,\ldots ,3$.
The
spatial symmetry group of the cube ${\cal G}({\cal C})$, (consisting of all
possible reflection and rotations which map the cube to itself) has a
structure
of the direct product
$${\cal G}({\cal C}) = C_2 \times S_4\eqno(2.2)$$
of the cyclic group of order $2$, $C_2$, generated by the central
inversion,
$P$ (which reverses the directions of all four diagonals $d_0,\ldots ,d_3$)
and
the symmetric group, $S_4$, of order $24$, consisting of the
transformations
which permute the diagonals preserving their
directions.\footnote{${}^2$}
{For the
even
permutation the corresponding transformation is a pure rotation of the
cube,
while for the odd permutation it is a rotation followed by the central
inversion, $P$.}

Let
$$s_{0123} = (s_0 \,\vert {\overline s}_1,\,{\overline s}_2,\,{\overline
s}_3\vert\,s_1,\,s_2,\,s_3\vert\,{\overline s}_0)\eqno(2.3)$$
denote the sequence of the spins $s_0,\ldots ,{\overline s}_0$,
corresponding
to their basic arrangement at the corners of $\cal C$ as shown in Fig.2.

Obviously, the transformations from ${\cal G}({\cal C})$ induce some
permutations
of the spins $s_0,\ldots ,{\overline s}_0$.  According to the above
discussion of
the structure (2.2) of ${\cal G}({\cal C})$ these transformations map the
sequence (2.3) either to $s_{ijk\ell}$ or to ${\overline s}_{ijk\ell}$,
$$\equalignno{s_{ijk\ell} &= (s_i\,\vert {\overline s}_j,\,{\overline
s}_k,\,{\overline s}_\ell \vert \,s_j,\,s_k,\,s_\ell\vert\,{\overline
s})&(2.4)\cr
\noalign{\vskip8pt}
{\overline s}_{ijk\ell} &= ({\overline s}_i\,\vert
s_j,\,s_k,\,s_\ell\vert\,{\overline s}_j,\,{\overline s}_k,\,{\overline
s}_\ell
\vert\,s)&(2.5)\cr}$$
where $(i,\,j,\,k,\,\ell)$ is some permutation of $(0,\,1,\,2,\,3)$.

Consider two elements of ${\cal G}({\cal C})$, specifying their action on
the
spins.  Let $(i,\,j,\,k,\,\ell)$ be any permutation of $(0,\,1,\,2,\,3)$,
and
let $R \in {\cal G}({\cal C})$,
$$R\,s_{ijk\ell} = {\overline s}_{\ell kij}\ \ ,\qquad R\,{\overline
s}_{ijk\ell}
= s_{\ell kij}\ \ ,\qquad R^4 = 1\eqno(2.6)$$
denote the $90^\circ$ rotation around the axes passing through the centers
of
the top and bottom faces of $\cal C$, while $T \in {\cal G}({\cal C})$,
$$T\,s_{ijk\ell} = s_{ikj\ell}\ \ ,\qquad T\,{\overline s}_{ijk\ell} =
{\overline s}_{ikj\ell}\ \ ,\qquad T^2 = 1\ \ ,\eqno(2.7)$$
denotes the reflection with respect to the plane passing through the
corners of
$\cal C$ occupied by the spins $s_0$, $s_3$, ${\overline s}_o$,
${\overline
s}_3$ in Fig.2.

One could easily check that these two elements $R$ and $T$ generate the
whole
group ${\cal G}({\cal C})$.  In particular, the central inversion, $P$,
$$P\,s_{ijk\ell} = {\overline s}_{ijk\ell}\ \ ,\qquad P\,{\overline
s}_{ijk\ell} = s_{ijk\ell}\ \ ,\qquad P^2 = 1\ \ ,\eqno(2.8)$$
can be expressed as
$$P = (R\,T)^3\ \ \ .\eqno(2.9)$$

\nt{2.2\qquad The angle parameterization}

\bigskip
Let $\theta_1$, $\theta_2$, $\theta_3$ denote angles of a spherical
triangle
and $a_1$, $a_2$, $a_3$ denote three sides of this triangle opposite to the
angles $\theta_1$, $\theta_2$, $\theta_3$.  Define the related variables
$$\equalign{\alpha_0 &= (\theta_1 + \theta_2 + \theta_3 - \pi)/2\ \ ,\qquad
\alpha_i = \theta_i - \alpha_0\ \ ,\cr
\noalign{\vskip8pt}
\beta_0 &= (2\pi - a_1 - a_2 - a_3)/2\ \ ,\qquad \beta_i = \pi - \beta_0 -
a_i\
\ ,\cr}\eqno(2.10)$$
for $i = 1,\,2,\,3$.  Choose $\theta_1$, $\theta_2$, $\theta_3$, so that
$\theta_1$, $\theta_2$, $\theta_3$, $\alpha_0,\ldots ,\alpha_3$,
$\beta_0,\ldots ,\beta_3$ are all real, between $0$ and $\pi$.	  Further,
define
(taking real positive values of roots)
$$\equalign{S_i &= \left[ \sin (\theta_i/2)\right]^{1/N}\ \ ,\quad C_i =
\left[
\cos (\theta_i/2)\right]^{1/N}\ \ ,\cr
\noalign{\vskip8pt}
 T_i &= \left[ \tan (\theta_i/2)\right]^{1/N}\ \ ,\quad z_i = \exp
(ia_i/N)\ \
,}\eqno(2.11)$$
for $i = 1,\,2,\,3$ and
$$u_i = \exp (i\beta_i/N)\ \ ,\quad c_i = \left[ \cos
(\alpha_i/2)\right]^{1/2}\ \ ,\quad i = 0,\ldots ,3\ \ \ .\eqno(2.12)$$
Now parameterize $p$, $p^{\prime}$, $q$, $q^{\prime}$ in (1.10) as
follows\footnote{${}^3$}{This parameterization has been obtained by a
formal generalization of eqs.(4.19),(4.24) of ref.[10] for
arbitrary values of $N$.}
$$p = z^{-1}_3\,T_1\ \ ,\quad p^{\prime} = \omega^{-1/2}z^{-1}_3T^{-1}_1\ \
,\quad q = T^{-1}_2\ \ ,\quad q^{\prime} = \omega^{-1/2}\,T_2,$$
then\footnote{${}^4$}{For $N = 2$ the variables $v_1,\ldots ,v_4$ here differ
 from those given by (7.19) of [18] merely by negating $v_3$ and $v_4$.}
$$\equalign{v_1 &= \omega^{-1} z_3 T_1 T_2\ \ ,\qquad v_2 = \omega^{-1/2}
z_3
T_2/T_1\ \ ,\cr
\noalign{\vskip8pt}
v_3 &= z^{-1}_3 T_1 T_2\ \ ,\qquad v_4 = \omega^{-1/2} z^{-1}_3 T_2/T_1\ \
\
.\cr}\eqno(2.13)$$
Now choose the phases of $\Delta(v_1),\ldots ,\Delta(v_4)$ such that
$$\equalign{\Delta (v_1) &= S_3/(C_1 C_2 u_3)\ \ ,\qquad \Delta (v_2) = C_3
v_1
/(S_1 C_2)\ \ ,\cr
\noalign{\vskip8pt}
\Delta (v_3) &= S_3 u_3/(C_1 C_2)\ \ ,\qquad \Delta (v_4) = C_3/(S_1 C_2
u_1)\
\ \ ,\cr}\eqno(2.14)$$
$$\Delta (v_4/v_3) = \Delta (v_2 /(\omega u_1)) = S^{-2}_1\eqno(2.15)$$

With these definitions the Boltzmann weight function (1.12) can be regarded
as a
function of the three independent variables $\theta_1$, $\theta_2$,
$\theta_3$
or, equivalently, of the four dependent variables $\alpha_0,\ldots
,\alpha_3$
constrained by the relation
$$\alpha_0 + \alpha_1 + \alpha_2 + \alpha_3 = \pi\eqno(2.16)$$
so we can write the LHS of (21) as $V(s_0,\ldots ,{\overline s}_0 \Vert
\alpha_0,\ldots ,\alpha_3)$.

Note, that the angles $\t_1,\t_2,\t_3$ can be viewed as the dihedral
angles between three ``rapidity planes'' passing through the cube
exactly as it is in the Zamolodchikov model.

\bigskip\bigskip
\nt{2.3\quad The normalization of the weights}

\bigskip
Now we like to fix the normalization factors $w(v,\,0)$ in (1.8) and $\rho$
in
(1.12).

First let us choose $w(v,\,0)$ such that
$$\prod^{N-1}_{a=0}\,w(v,\,a) = 1\ \ \ .\eqno(2.17)$$
With this normalization define
$$\equalign{D_+(v) &= (\det\negthinspace_N \Vert w(v,\,a - b)\Vert)^{1/N}\
\
\ ,\cr
\noalign{\vskip8pt}
D_-(v) &= (\det\negthinspace_N\Vert 1/w (v,\,a -
b)\Vert)^{1/N}\ \ \ .\cr}\eqno(2.18)$$
Also, set
$$\equalign{S_+ &= (\det\negthinspace_N \Vert \g(a,\,b)\Vert)^{1/N}\ \ \
,\cr
\noalign{\vskip8pt}
S_- &= (\det\negthinspace_N \Vert 1/\g(a,\,b)\Vert)^{1/N}\ \ \
.\cr}\eqno(2.19)$$
Regarding $a$, $b$ in (1.19), (1.20) as matrix indices running the values
$0,\ldots ,N-1$, and taking the determinants of the both sides of these
equations
one gets by using (2.17)--(2.19)
$$\equalign{\varphi_1(v_1,\,v_2) &= D_+ (v_2)S_-/D_-(v^{\prime}_1)\ \ \
,\cr
\noalign{\vskip8pt}
\varphi_2 (v_3,\,v_4) &= D_-(v_4)S_+/D_+(v^{\prime}_3)\ \ \
,\cr}\eqno(2.20)$$
where $v^{\prime}_1$, $v^{\prime}_3$ are given by (1.21).  Explicit
calculations
with the equations (1.8) give
$$\equalign{D_{\pm}(v) &= c_{\pm}\ (v/\Delta(v))^{(N-1)/2}\cr
\noalign{\vskip8pt}  S_+ S_- &= N\cr}\eqno(2.21)$$ where $c_{\pm}$ are
(inessential)
constants.
Note, in particular, that when $v_1,\ldots,v_4$ are parameterized by
(2.13)-(2.15) we have
$$
\phi_1(v_1,v_2)\phi_2(v_3,v_4)=N\left({\sin\t_2\over\sin\t_3}\right)^{(N-1)/N}\eqno(2.22)
$$

Further, set the normalization factor in (1.12) as
$$\equalignno{\rho &= (2\xi)^{2(N-1)/N}/N\ \ \ ,&(2.23)\cr
\noalign{\vskip8pt}
2\xi &= \left(\h \sin \theta_3\right)^{1/2}/(c_0 c_1 c_2 c_3)&(2.24)\cr}$$
where $c_0,\ldots,c_3$ are given by (2.12).

We want to calculate the free energy, or equivalently the partition
function
per site
$$\kappa = Z^{1/M}\ \ \ .\eqno(2.25)$$

Note that when $\alpha_0 = \alpha _2 = 0$ we have from (2.10) - (2.13)
$$v_1 = v_2 = v_3 = v_4 = 0\ \ \ ,\qquad 2\xi = 1\ \ ,\eqno(2.26)$$
while from (1.8)
$$w(0,\,a) = w(0,\,0)\ \ \ ,\quad \forall a\ \ \ .\eqno(2.27)$$
Substituting (2.26), (2.27) into (1.12) and using (1.15), (1.16) one gets
$$V = \delta_{s_0-{\overline s}_2,s_2 - {\overline s}_0}\ \ \
.\eqno(2.28)$$
Ignoring irrelevant boundary contributions, it follows that for $M$ large
$$\kappa = 1\ \ \ ,\quad {\rm when}\ \ \alpha_0 = \alpha_2 = 0\ \ \
.\eqno(2.29)$$

\bigskip\bigskip
\nt{2.4\quad The symmetry properties of the Boltzmann weights}

\bigskip

The definition (1.12) of the \BW\ function $V$ is obviously rather
asymmetric  with respect to the orientation of the elementary cube
(see Fig.3). We shall see, however,  that despite this visual
asymmetry, the \BW\ function (1.12) has a remarkable hidden symmetry.
Below we shall show that up to the equivalence transformation factors,
which do not affect the partition function, the \BW\ function
remains unchanged upon the permutations the corner spins $s_0,\ldots,
\overline{s}_0$ induced by symmetry transformations of the cube
complemented by corresponding transformation of the variables
$v_1,v_2,v_3,v_4$ (or, conveniently,  the variables
$\a_0,\ldots,\a_3$). Obviously, it is enough to prove this just for
two transformations (2.6) and (2.7) since they generate the whole
symmetry group of the cube.

 From (1.3),(1.18), (2.17) it follows that
$$
w(v,0) w(\tilde{v},0) = e^{i \pi (N^2-1)/6}\eqno(2.30)
$$
 is a constant independent of $v$. Let us of
interchange  $\a_1$ with $\a_2$ leaving $\a_0,\a_3$ intact. From
(2.13) this is equivalent to the replacement of $v_1,v_2,v_3,v_4$ by
$v_1,\tilde{v}_4,v_3,\tilde{v}_2$ respectively. Using (2.30) and noting
that
(2.23) remains unchanged  one can easily trace the effect of
this interchange on the weight function (1.12)
$$
V(s_{0123}||\a_0,\a_1,\a_2,\a_3)={\Phi(s_2-\overline{s}_1)\over
\Phi(s_1-\overline{s}_2)}V(s_{0213}||\a_0,\a_2,\a_1,\a_3)\eqno(2.31)
$$
where we have used the short notations (2.4) for the spin
configuration of the corner spins. This gives the transformation law
of $V$ under the reflection $T$, (2.7).

Further, using (1.15), (1.16) rewrite (1.12) in the form
$$\eqalign{
{ s(s_0-\overline{s}_3,s_0-\overline{s}_2) \rho\over
\,\,\, s(\overline{s}_0-s_3,\overline{s}_0-s_2) N}& \sum_{\mu=0}^{N-1}\biggl\{
{s(s_3+\overline{s}_3,\mu)\over  s(s_0+\overline{s}_0,\mu)}\times\cr
F_1(v_1,v_2|s_0+s_1-\overline{s}_2-\overline{s}_3,\overline{s}_2-s_0+\mu)
&F_2(v_3,v_4|\overline{s}_0+\overline{s}_1-s_2-s_3,s_2-\overline{s}_0-\mu)
\biggr\}\cr
}\eqno(2.32)
$$
Applying now (1.19), (1.20) and taking into account (2.14), (2.15),
(2.20)-(2.24) one obtains
$$\eqalign{
V(s_{0123}||\a_0,\a_1,\a_2,\a_3)=&\cr
{s(s_0-\overline{s}_3,s_0-\overline{s}_2)
\over s(\overline{s}_0-s_3,\overline{s}_0-s_2)}
&{w(v_4/v_3,s_0+s_1-\overline{s}_2-\overline{s}_3)\over
w(v_4/v_3,\overline{s}_0+\overline{s}_1-s_2-s_3)}
V(\overline{s}_{3201}||\a_3,\a_2,\a_0,\a_1)
}\eqno(2.33)$$
which gives the required transformation of $V$ under the rotation,
$R$, (2.6). Combining now eqs.(2.32) and (2.33) and using (1.18),
(2.30) for the functions $w$ in (2.33) one gets the relation
corresponding to the  $RT$-transformation
$$\eqalign{
V(s_{0123}||\a_0,\a_1,\a_2,\a_3)=&\cr
{\Phi(s_0)\Phi(s_3)\over
\Phi(\overline{s}_0)\Phi(\overline{s}_3)}
{s(\overline{s}_2,\overline{s}_0-s_3)
\over s(s_2,s_0-\overline{s}_3)}
&{w(v_2v_3,s_1+s_3-\overline{s}_0-\overline{s}_2)\over
w(v_2v_3,\overline{s}_1+\overline{s}_3-s_0-s_2)}
V(\overline{s}_{3102}||\a_3,\a_1,\a_0,\a_2)
}\eqno(2.34)
$$
where the argument of $w$'s can be written as
$$
v_2v_3=q'/q=\w^{-1/2}T_2^2.\eqno(2.35)
$$
Remembering eq.(2.9) and iterating  (2.34) three times one obtains
$$\eqalign{
&{V(s_{0123}||\a_0,\a_1,\a_2,\a_3)\over
V(\overline{s}_{0123}||\a_0,\a_1,\a_2,\a_3)}=
{\Phi^2(s_0)\over
\Phi^2(\overline{s}_0)}
{s(\overline{s}_2,\overline{s}_0-s_3)
\over s(s_2,s_0-\overline{s}_3)}
{s(s_0,s_3-\overline{s}_2)
\over s(\overline{s}_0,\overline{s}_3-s_2)}
{s(\overline{s}_3,\overline{s}_2-s_0)
\over s(s_3,s_2-\overline{s}_0)}\cr
&{w(v_2v_3,s_1+s_3-\overline{s}_0-\overline{s}_2)\over
w(v_2v_3,\overline{s}_1+\overline{s}_3-s_0-s_2)}
{w(v_5,\overline{s}_1+\overline{s}_2-s_0-s_3)\over
w(v_5,s_1+s_2-\overline{s}_0-\overline{s}_3)}
{w(v_4/v_3,s_0+s_1-\overline{s}_2-\overline{s}_3)\over
w(v_4/v_3,\overline{s}_0+\overline{s}_1-s_2-s_3)}
}\eqno(2.36)
$$
where
$$
v_5={v_4 \D{v_1} \D{v_3}\over v_3 \D{v_2} \D{v_4}}=\w^{-1/2}T_3^2,\qquad
\D{v_5}=C_3^2.
\eqno(2.37)
$$
One can check that (2.36) is exactly the three-dimen\-sional
star-star relation conjectured previously (eq.(6.1) of ref.[10]).

As we remarked before the angles $\t_1,\t_2,\t_3$ can be interpreted
as the dihedral angles between the three rapidity planes rigidly
connected with the cube. Then, from geometric considerations these
angles should be very simply transformed by
the cube symmetry group ${\cal G(C)}$. Namely, the related variables
$\a_0,\a_1,\a_2,\a_3$  given by (2.10) should just permute for any
transformation from ${\cal G(C)}$. This is entirely consistent with
(2.31), (2.33).

\beginsection{3.  Partition function}

\bigskip

\nt 3.2\quad Factorization and commutativity.
\bigskip

Consider two successive layers of $\L$ with $l$ spins $\f$ on the lower
layer, $\f'$ on the upper. In the center of each intervening cube we
have  the central spin $\s$ as in Fig. 3. Let $\f''$
denote the set of all these $\s$-spins between $\f$ and $\f''$.
Then because  the top spins $s_0,s_1,\os_2,\os_3$ in Fig. 3 interact
only with one another and with $\s$, and similarity for thee bottom
spins, we can write the elements of the transfer matrix $T$ as
$$
T_{\f,\f'} = \rho^l\sum_{\f''}X_{\f,\f''}Y_{\f'',\f'} , \eqno{(3.1)}
$$
where
$$
\eqalign{
X_{\f,\f''}&=\prod_{cubes}{w(v_3,s_3-\os_0-\s)\g(\os_0,\s)
\over w(v_4,\os_1-s_2-\s)
\g(s_2,\s)},\cr
Y_{\f'',\f}&=\prod_{cubes}{w(v_2,s_1-\os_2+\s)\g(\os_0,\s)
\over w(v_1,\os_3-s_0+\s)\g(s_2,\s)}.\cr
}\eqno(3.2)
$$
The products are over the all the $l$ cubes between the two layers;
for each cube $s_0,s_1,s_2,s_3$, $\os_1,\os_2,\os_3,\os_0$ are the eight
spins shown in Fig 3.

Obviously we can regard $X_{\f,\f''}$ as the element of a matrix $X$.
 From (3.2) this matrix depends on $v_3$ and $v_4$, so  it can be
written as $X(v_3,v_4)$. With similar conventions for $Y$, (3.1)
implies
$$
T=\rho^lX(v_3,v_4)Y(v_1,v_2).\eqno (3.3)
$$
Thus $\rho^{-l}T$ factors into a product of two matrices, one
dependent on $\t_1,\t_2,\t_3$ only via $v_3$ and $v_4$, the other via
$v_1$ and $v_2$.

In [10] it was shown that transfer matrices $T$ form  two-parameter
commuting families. More precisely, when parameterizing $p,p',q,q'$
 through the angles (2.13), the eqs. (1.5), (2.11), (2.17), (2.22)
of ref. [10] imply that two transfer matrices (with different values
of $\t_1,\t_2,\t_3$ ) commute provided they have the same value of
$\t_1$
$$
[T(\t_1,\t_2,\t_3),T(\t_1,\t_2',\t_3')]=0, \qquad \forall
\t_2,\t_3,\t_2',\t_3'   \eqno (3.4)
$$
The proof was based on the Yang-Baxter equation for the
$sl(n)$-chiral Potts model [11, 17,19]. Alternatively we can establish
the commutativity property (3.4) directly from the
3-dimensional star-star relation (2.36) not referring to  those
results.
 This will be done below in this section.

If $\t_1$ is known then $v_2$ can be determined from $v_1$ and $v_4$
from $v_3$ by  corollaries of (2.14).
$$
\eqalign{
v_2&= \w^{1/2}\left(\cot{\t_1\over 2}\right)^{2/N}v_1,\cr
v_4&= \w^{-1/2}\left(\cot{\t_1\over 2}\right)^{2/N}v_3.\cr
}\eqno(3.5)
$$
The text step is to show that the factorization property (3.3) remains
true when the matrices $T,X,Y$ are appropriately diagonalized.
To ensure this, it is necessary to slightly modify the model.

\noindent
There are two sorts of
vertical faces in $\L$: those whose perpendiculars run in front-to-back
direction
(such as $s_0 \os_2 s_3 \os_1$ and $\os_3 s_1 \os_0 s_2$
in Fig.~3), and those whose perpendiculars run
right-to-left. Call the former type ``FB'', the
latter ``RL''. At the center of each FB face place a spin $\mu$, with
values $0,\ldots,N-1$. Let the spins on the front and back faces in Fig.3
be $\mu$ and $\mu'$, respectively. Choose them  so that
$$
\s=\mu-\mu'\pmod{N}. \eqno(3.6)
$$
Do this for all cubes in $\L$ . If $\s'$ is the spin behind $\s$, and
$\s''$ is the spin behind that, etc.; then on using the cyclic
boundary conditions we observe that
$$
\s+\s'+\s''+\cdots=(\mu-\mu')+(\mu'-\mu'')+(\mu''-\mu''')+\cdots=0.\pmod{N}
\eqno(3.7)
$$
(Each $\mu$-spin occurs twice with the opposite signs. If $\L$ has $n$
layers perpendicular to the front-to-back direction then there are $n$
$\s$-spins on the LHS of (3.6)). Thus we can
use (3.6) only if the sum of each horizontal front-to-back line of
$\s$-spins is constrained to be zero. This is merely  a change of
boundary conditions, and in the limit of $n$ large it should
have no effect on the partition function per site $\kappa$.
We shall refer to
the model subject to these constraints as the ``modified model''.

For the modified model (3.1) and (3.2) formally remains the same, but $\phi''$
is now the set of all $\mu$-spins and $\s$ in (3.2) is now given by
(3.6). Note also that in this case the matrices $X,Y,T$ are unchanged
under overall shifts of all $\m$-spins, or all $s$-spins, on any
 front-to-back line.
Therefore $X,Y$ and $T$ have non-zero entries only in the diagonal block with
respect to the subspace invariant under all such shifts.

For the moment, let us ignore the constraints (3.5), and work with
the original model. The elements of the transfer matrix $XY$ in
(3.1)-(3.3) are obtained by taking the product over all cubes in
 a layer of the function $V(s_{0123}
\parallel \a_0,\dots,\a_3)$ given by (1.12), (2.3). Let us replace
the function $V$ by $\overline V = V(\os_{0123}\parallel
\a_0,\dots,\a_3)$ with $\overline s_{0123}$ given by (2.5)
 and denote the corresponding transfer matrix as $\overline T$.
Then with the same arguments which led  to (3.3) one can show that
$$
\overline T = \rho^l\hat X(v_1,v_2)\hat Y(v_3,v_4),\eqno(3.8)
$$
where $\hat X$, $\hat Y$ are similar but not identical to $X$, $Y$ in
(3.3).
 From (2.36) the functions $V$ and $\overline V$ differ by the
equivalence  transformation factors only. If we take the product of
these factors over all cubes in a layer then the result can be put
in the form $L(\phi)/L(\phi')$, where $\phi$ is the set of spins
on the lower layer, $\phi'$ on the upper.

We have therefore shown that
$$
[\hat X(v_1,v_2)\hat Y(v_3,v_4)]_{\phi,\phi'}=
L(\phi)[ X(v_3,v_4) Y(v_1,v_2)]_{\phi,\phi'}/L(\phi')\eqno(3.9)
$$
or in matrix notation
$$
\hat X(v_1,v_2)\hat Y(v_3,v_4)=L X(v_3,v_4) Y(v_1,v_2)L^{-1}
\eqno(3.10)
$$
where $L$ is the diagonal matrix with elements $L(\phi)\d(\phi,\phi')$.
It depends on $\t_1$, but not on any other parameters.

Now introduce the constraints (3.6). This is equivalent to introducing
a factor
$$
\prod\left\{{1\over N}\sum_{k=0}^{N-1}\w^{k(\s+\s'+\s''+\dots)}\right\}
\eqno(3.11)
$$
into summand in (3.1), where $\s,\s',\s'',\dots$ are the $n$ center
spins
an a horizontal front-to-back line, and the outer product is over
$l/n$ such lines in the horizontal layer of spins $\phi''$.

Let $V_k$ be a function that differs from $V$ only in that an extra
single spin factor $\w^{k\s}$ is put into the summand in (1.12).
Inserting (3.11) into (3.1) we obtain
$$
T = \prod\left\{{1\over N}\sum_{k=0}^{N-1}\prod V_k\right\}\eqno(3.12)
$$
where the inner product is over $n$ cubes in a horizontal
front-to-back line and again the outer product is over all $l/n$
such lines in a layer. From (1.2)
$$
V_k(s_0|\overline s_1,\overline s_2,\overline
s_3|s_1,s_2,s_3|\overline s_0)=V(s_0+k|\overline s_1,\overline s_2,
\overline s_3+k|s_1,s_2,s_3|\overline s_0)
\eqno(3.13)
$$
Shifting $s_0$ and $\overline s_3$ in (2.36) we obtain
$$
{V_k\over \overline V_k}= {\g(k,s_0-\overline s_3)\over \g(k,\overline
s_0-s_3)}
{w(v_5,\overline s_1+s_2-s_0-s_3-k)\over
w(v_5,s_1+s_2-\overline s_0-\overline s_3-k)}
{w(v_5,s_1+s_2-\overline s_0-\overline s_3)\over
w(v_5,\overline s_1+\overline s_2-s_0-s_3)}
{V\over \overline V}
\eqno(3.14)
$$
where $\overline V_k =V_k(\overline s_{0123})$. If we take the product
of each side of (3.14) over the $n$ cubes in a horizontal
front-to-back line, the $w$ and $s$ factors cancel.
It follows that we still have relations (3.10) in the modified
model\footnote{${}^5$}{Applying similar arguments to the product of the
weight function $V$ along one front-to-back line of the cube,
rather that to the whole horizontal layer, one obtains precisely the
two dimensional star-star relation of the $sl(n)$-chiral Potts model
(eq. (3.19) of [11]).}.

Now consider the case when $v_2=v_3=1$. From (3.5) it follows then
$$
v_1={1\over\w v_4}=\w^{-1/2}(\tan {\t_1\over 2})^2/N      \eqno(3.15)
$$
while (1.12) gives
$$
V(s_0,\dots ,\overline s_0)=
{\lambda_1\delta(\overline s_2-s_1,s_3-\overline s_0)\over
w(v_1,\overline s_3-s_0-s_1+s_2)w(v_4,\overline s_1-s_2-s_3-\overline
s_0)}
\eqno(3.16)
$$
where $\lambda_1$ is a scalar factor. (From (2.17) $\lambda_1$ is
infinite, but this is just a feature of the normalization (2.17),
and can readily be removed, leaving the following arguments intact).
The function $V_k$ contains extra $\w^{k(s_0-\overline s_3)}$ factor
but this cancel out of the product in (3.12), so in this case
$$
T_{\phi,\phi'}=\prod_{cubes}V(s_0,\dots,\overline s_0)\eqno(3.17)
$$
where $V$ is given by  (3.16) and the product is over all cubes in a
layer. Substituting (3.16) into (3.17) and taking into account (1.18),
(2.30) one obtains
$$
T_{\phi,\phi'}=\lambda_2\prod_{cubes}\delta
(\overline s_2-s_1,s_3-\overline s_0)  \eqno(3.18)
$$
where $\lambda_2$ is a constant.

The product of Kronecker delta-functions is zero unless $\phi  =
\phi'$, or if $\phi,\phi'$ differ only by overall shifts of spins on
some horizontal front-to-back lines. Restricting attention to the
subspace invariant under these shifts we then obtain from (3.3),
(3.18)
$$
X(1,v_4)Y(v_1,1)=\lambda I\eqno (3.19)
$$
where $\lambda$ is another (nonzero) constant, $I$ is the unit matrix
and $v_1$ and $v_4$ are related by (3.15). For the modified model the
matrices in (3.18) are square and hence non-singular and
invertible\footnote{${}^6$}{Curiously enough, it is this apparently
innocuous statement that fails for the original model and is the
reason for introducing the constraints (3.6).}.

Regard $\t_1$ as fixed, $v_1$ and $v_3$ as independent variables, and
$v_2$ and $v_4$ as given by (3.5). Then we can suppress $\t_1$, $v_2$,
$v_4$ dependence and write (3.10) as
$$
\hat X(v_1)\hat Y(v_3)=LX(v_3)Y(v_1)L^{-1}\eqno(3.20)
$$
Then (3.19) gives
$$
X(1)Y(x)=\hat X(x)\hat Y(1)=\lambda I\eqno (3.21)
$$
where $x=\w^{-1/2}(\tan {\t_1/2})^{2/N}$ is the value of $v_1$
when $v_2=1$.

Using (3.20), (3.21) we can now perform the following transformations
$$\eqalign{
X(v_3)Y(x)X(1)Y(v_1)=\l X(v_3) Y(v_1) =\l L^{-1} \hat X(v_1)\hat Y(v_3)L&=\cr
L^{-1} \hat X(v_1)\hat Y(1)\hat X(x)L\hat
Y(v_3)L=X(1)Y(v_1)X(v_3)Y(x)&\cr
}\eqno(3.22)
$$
Hence the matrix products $X(1)Y(v_1)$ and $X(v_3)Y(x)$ commute for
all values of $v_1$ and $ v_3$. Assuming that these commuting matrix
products can be simultaneously diagonalized it follows that there must
exists diagonal matrices $A(v_3)$ and $B(v_1)$ and a non-singular
matrix $P$ (independent of the variables $v_1$ and $v_3$) such that
$$
X(v_3)Y(x)=PA(v_3)P^{-1},\qquad
\l^{-1}X(1)Y(v_1)=PB(v_1)P^{-1}\eqno(3.23)
$$
Setting $Q=Y(x)P$ and remembering that all matrices depend implicitly
on $\t_1$, we finally obtain
$$\eqalign{
X(v_3,v_4)=&P(\t_1)A(v_3,v_4)Q^{-1}(\t_1)\cr
Y(v_1,v_2)=&Q(\t_1)B(v_1,v_2)P^{-1}(\t_1)\cr
}\eqno(3.24[B)
$$
Here all the matrices depend on $\t_1,\t_2,\t_3$ only via the
arguments explicitly shown.

Note that (3.23) imply the commutativity relation (3.4) for the
modified model. (Using then simple arguments like those at the
end of Sect.2 of ref.[10] we can easily extend (3.4) to the original
model as well).

Thus, the relations (1.15)-(1.20) imply the star-star relation (2.36)
which in its turn imply (3.20) and the relation (3.24).

 From (1.2),(3.3) and (3.24) it follows that
$$\eqalign{
Z=&\rho^M \trace \left(A(v_3,v_4) B(v_1,v_2)\right)^m\cr
=&\rho^M \sum_j(A_{jj}(v_3,v_4) B_{jj}(v_1,v_2))^m\cr}
$$
where the $j$-summation is over all the diagonal elements $A_{jj}$
and $B_{jj}$ of $A$ and $B$. Assuming the largest term in the
summation is unique and writing $a^l$ for $A_{00} $ and $b^l$ for $B_{00} $
(choosing $j=0$ for the largest term), it follows that from (3.25),
(2.25)
$$
\kappa=\rho\, a(v_3,v_4) b(v_1,v_2), \eqno(3.25)
$$
where $ \rho$ is given by (2.23).
This is the factorization property which will be used in the next Section.
\bigskip

\noindent 3.2 Partition function per site.
\bigskip

If $\L$ is infinite in all directions, we can evaluate the partition
function per site solely from the symmetry and the factorization
properties. In fact, the definition (1.1) is invariant under the
symmetry transformation of the lattice generated by the symmetry group
of the cube $\cal G(C)$. On the other hand from (2.31) and (2.33) the
effect of any such transformation on the weight function $V$ is to
multiply it by some equivalence transformation factors and permute
$\a_0, \a_1,\a_2,\a_3$. Since two such transformations in  (2.31),
(2.33) generate all possible permutations of $\a_0, \a_1,\a_2,\a_3$,
the partition function obey the following symmetry relation
$$
\kappa(\a_0, \a_1,\a_2,\a_3)= \kappa(\a_i, \a_j,\a_k,\a_l)\eqno(3.26)
$$
where $(i,j,k,l)$ is any permutation of $(0,1,2,3)$.

Thus we have two functional equations for $\kappa$, the factorization
property (3.25), (2.23) and the symmetry property (3.26). We shall see
that these two functional equations define $\kappa$ up to
multiplicative constant which is determined then from the
normalization (2.29).

The \PFS\ $\kappa$ obviously depends on the number of the spin states
$N$. We  can exhibit this writing it as $\kappa_N$. Let  us
concentrate on the $N$-dependence in the functional equations. The
symmetry relation (3.26) does not involve $N$ at all. However, there
are two possible sources of the $N$-dependence in the factorization
equation (3.25). First $\rho$ in (3.25) depends on $N$ as given by
(2.23). Second, the formulae (2.13) expressing the variables
$v_1,\ldots,v_4$ through the angles $\t_1,\t_2,\t_3$ have the N-dependence
coming from (2.11). This, however, can be readily removed since the
$N$-th powers of the variables $v_1,\ldots,v_4$ are the same for any
$N$ (the $N$-th powers of the RHS's of (2.13) do not have any
$N$-dependence). So redefining the functions $a$ and $b$ we we can
rewrite (2.25) in the form
$$
\kappa=\rho \, \overline{a}(x_3,x_4)\overline{b}(x_1,x_2)\eqno(3.27)
$$
where $x_i=v_i^N$, $i=1,\ldots,4$,
$$\eqalign{
x_1=e^{ia_3} \tan(\t_1/2)\tan(\t_2/2),\qquad
&x_2=-e^{ia_3} \cot(\t_1/2)\tan(\t_2/2) \cr
x_3=e^{-ia_3} \tan(\t_1/2)\tan(\t_2/2),\qquad
&x_4=-e^{-ia_3} \cot(\t_1/2)\tan(\t_2/2) \cr
}\eqno(3.28)
$$
are expressed merely through the angles $\t_1,\t_2,\t_3$ and the only
$N$-dependence in the RHS of (3.27) comes from the factor $\rho$.
Writing $\kappa_N$ in the form
$$
\kappa_N=\lambda_N (\psi)^{2(N-1)/N} \eqno(3.29)
$$
where $\lambda_N$ is a numerical constant and substituting it into
(3.27) we get the factorization equation for $\psi$
$$
\psi=(2\xi) \overline{\overline{a}}(x_3,x_4)
\overline{\overline{b}}(x_1,x_2)\eqno(3.30)
$$
where $\xi$ is  given by (2.24). The symmetry relation for $\psi$
remains, clearly, the same as (3.26). Comparing (3.29) with eq.(7.33)
of ref.[18] we see that  $\psi$ satisfies the same factorization
and symmetry relation as those for the \PFS\ of the \ZM,  $\kappa_2$
(which is the $N=2$ case of the considered model). It was shown in [18]
that these relations determine   $\kappa_2$ to within a numerical
constant,
so we can conclude that
$$
 \psi=\lambda \kappa_2\eqno(3.31)
$$
If we define the function $G(\b)$ by
$$
\log G(\b)={1\over2\pi}\int_0^\b \left[x\cot x
+{\pi\over2}\tan{x\over2}-\log(2\sin x)\right] dx \eqno(3.32)
$$
then the result of [18] can be written as
$$
\kappa_2={1\over2} G(\b_0)G(\b_1)G(\b_2)G(\b_3)\eqno(3.33)
$$
where $\b_0,\b_1,\b_2,\b_3$ are given by (2.11).
Note that (3.33) satisfies the normalization (2.29).
Taking this into account and substituting (3.31) into (3.29) and then
into (2.29) we obtain
$$
\log \kappa_N(\t_1,\t_2,\t_3)={2(N-1)\over
N}\log\kappa_2(\t_1,\t_2,\t_3)  \eqno (3.34)
$$
which seems to be very simple and interesting result.

\beginsection{4. Free Boson Model}

It is interesting to learn whether the method described above is unique to
the considered model or it can be applied to the other models as well.
In the other words how many solutions are there for the relations
(1.15)--(1.20)? At the moment we do not  know  an answer to this question but
can give one more example. It is the  free boson (Gaussian) model on
the three dimensional lattice.

Let now the spins  at the sites of the lattice take continuous real
values $-\infty<s<\infty$ and the summation over each spin $s$ is
replaced by the integration $\int_{-\infty}^{\infty}\ldots ds$. With
these modifications the definitions (1.1), (1.12) remain valid, but
the
weights $w$ and $\g$ in (1.12) are now given by
$$
w(v,a)=\exp\left({-i a^2 v\over 2(1-v)}\right),\quad
\g(a,b)=\exp(-i ab),\quad
\Phi(a)=\exp(-i a^2/2).\eqno(4.1)
$$
Obviously,
$$
w(v,0)=1.\eqno(4.2)
$$
If we define $\tilde{v}$ and $ \Delta(v)$ as
$$
\tilde{v}=1/v, \qquad \Delta(v)=1-v,\eqno(4.3)
$$
for any $v$ then the relations (1.15), (1.17), (1.18) remain valid
while (1.16) is replaced with
$$
\int_{-\infty}^{\infty} \g(a,b) \g(-b,c) db= 2\pi \delta(a-c).\eqno(4.4)
$$
Further, if we also replace the spin summation
by the integration in (1.19)--(1.21) then these formulae become valid
as well (provided one formally sets $N=1$ in (1.4), (1.6), i.e.
$\Delta(v)$ is now given by (4.3) and $\w=1$). The scalar functions
$\phi_1$
and $\phi_2$ are now given by
$$
\eqalign{\phi_1(v_1,v_2)&=\left[{v_2-v_1\over2\pi i
(1-v_1)(1-v_2)}\right]^{-1/2}
,\cr
\phi_2(v_3,v_4)&=\left[{v_3-v_4\over2\pi i
(1-v_3)(1-v_4)}\right]^{-1/2}
.\cr
}\eqno(4.5)
$$
Let us parameterize $v_1,\ldots,v_4$ through the angles
$\t_1,\t_2,\t_3$ setting $v_i=x_i$, $i=1,\ldots,4$, where $x$'s are
given by (3.28). Then, in particular, we have (cf. (2.22))
$$
\phi_1(v_1,v_2)\phi_2(v_3,v_4)=2\pi \left(
{\sin\t_2\over\sin\t_3}\right)^{-1}.
\eqno(4.6)
$$
Choose the normalization factor $\rho$ in (1.12) as (cf. (2.23))
$$
\rho=(2\xi)^{-2}/(2\pi),\eqno(4.7)
$$
where $\xi$ is given by (2.24).
Finally the RHS of (2.28) is replaced by $\delta(s_0-s_2-\os_2+\os_0)$
and we still have (2.26).

We have now  all the required relations to reproduce the
results(3.26), (3.27) for the free boson model and calculate its
partition function per site $\kappa_B$. In fact all the calculations
are identical to those we did before except that the exponents in (4.7) and
(2.23) are different and the factors $N$  in (1.16) and (2.22) are
replaced by $2\pi$ in (4.4), (4.6). (This factors are mutually cancel
and therefore have no effect on the final result). Taking into account
these minor differences we obtain
$$
\log \kappa_B=-2 \log \kappa_2\eqno(4.8)
$$
A nice feature of the Gaussian model is that its partition function can
be found by the elementary ``Gaussian integration'' directly from the
definition (1.1). Together with (3.33) this means that the partition function
per site of the $N$-state model $\kappa_N$ can be expressed as a
{\it rational} power of a  free boson determinant. We see no obvious
reasons of this fact and it  would be quite
interesting to learn why it is so. We postpone the evaluation and the
analysis of this determinant to the separate publication.

\bigskip
\bigskip
{\it Acknowledgment}. We  are indebted to Professor L.G.Kov\'acs
for very useful discussions. When the
manuscript of this paper has been in preparation we received the
preprint [20] partially overlapping with our Sect.2. The authors of
that paper found a neat form of the symmetry relations for the
Boltzmann weight function of the elementary cube absorbing all the
equivalence transformation factors into the definition of the weights.
(However, they do not calculate the normalization factors in these relations).
Moreover that paper contains a discussion of the tetrahedron relation
in the model. We thank the authors of [20] for sending the preprint of their
paper.

\beginsection          References

\item{[1]} L. Onsager, Phys. Rev. 65 (1944) 117.

\item{[2]} C.N. Yang, Phys. Rev. Lett. (1967) 1312.

\item{[3]} R.J. Baxter, Ann. Phys. 70 (1972) 193.

\item{[4]} L.D.Faddeev, Sov. Sci. Rev. C1 (1980) 107--155.

\item{[5]} R.J.Baxter, {\sl Exactly Solved Models in Statistical Mechanics}
 (Academic Press, London, 1982).

\item{[6]} A.B.Zamolodchikov, Zh. Eksp. Teor. Fiz. 79 (1980) 641--664
[English trans.: JETP 52 (1980) 325--336];\hfill\break
A.B.Zamolodchikov, Commun. Math. Phys. 79 (1981) 489--505.

\item{[7]} V.V. Bazhanov, Yu.G. Stroganov, Teor. Mat. Fiz.
  52 (1982) 105--113 [English trans.: Theor. Math. Phys. 52 (1982) 685--691].

\item{[8]} M.T.Jaekel, J.M.Maillard, J. Phys. A15 (1982) 1309.

\item{[9]} R.J.Baxter, Commun. Math. Phys. 88 (1983) 185.

\item{[10]} V.V. Bazhanov, R.J.Baxter, J. Stat. Phys. 69 (1992) 453-485.

\item{[11]}
V.V. Bazhanov, R.M. Kashaev, V.V. Mangazeev, Yu.G. Stroganov,
Commun. Math. Phys. 138 (1991) 393--408.

\item{[12]}
H. Au-Yang, B.M. McCoy, J.H.H. Perk, S. Tang, M. Yan, Phys. Lett.
A123 (1987) 219.

\item{[13]}
B.M. McCoy, J.H.H. Perk, S. Tang, C.H. Sah, Phys. Lett. A125 (1987) 9.

\item{[14]}
R.J. Baxter, J.H.H. Perk and H. Au-Yang, Phys. Lett. A128 (1988) 138.

\item{[15]}
V.V. Bazhanov, Yu.G. Stroganov, J. Stat. Phys. 59 (1990) 799.

\item{[16]}
E. Date, M. Jimbo, K. Miki, T. Miwa,  Phys. Lett. A 148 (1990) 45.

\item{[17]}
E. Date, M. Jimbo, K. Miki, T. Miwa,  Commun. Math. Phys. 137
(1991) 133.

\item{[18]} R.J.Baxter, Physica 18D (1986) 321--347.

\item{[19]}
R.M. Kashaev, V.V. Mangazeev, T. Nakanishi, Nucl. Phys. B362 (1991) 563.

\item{[20]} R.M. Kashaev, V.V. Mangazeev, Yu.G. Stroganov, ``Spatial Symmetry
Local Integrability and Tetrahedron Equations in the Baxter-Bazhanov
Model''. Preprint IHEP-92-62, Serpukhov, 1992.

\bigskip
\centerline{\bf Figure Captions}
\bigskip

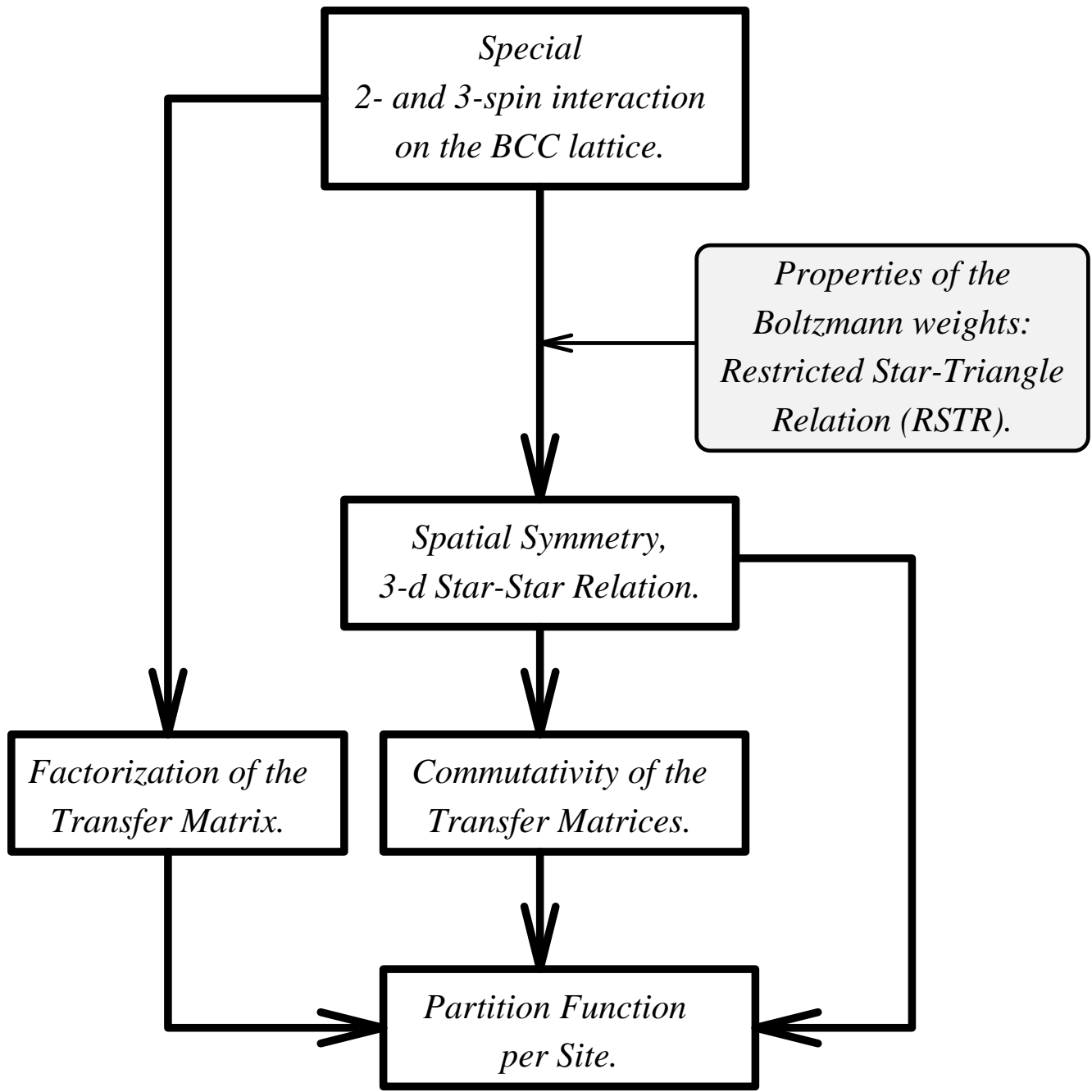
\bigskip Fig.1. This figure explains the approach used in this paper.

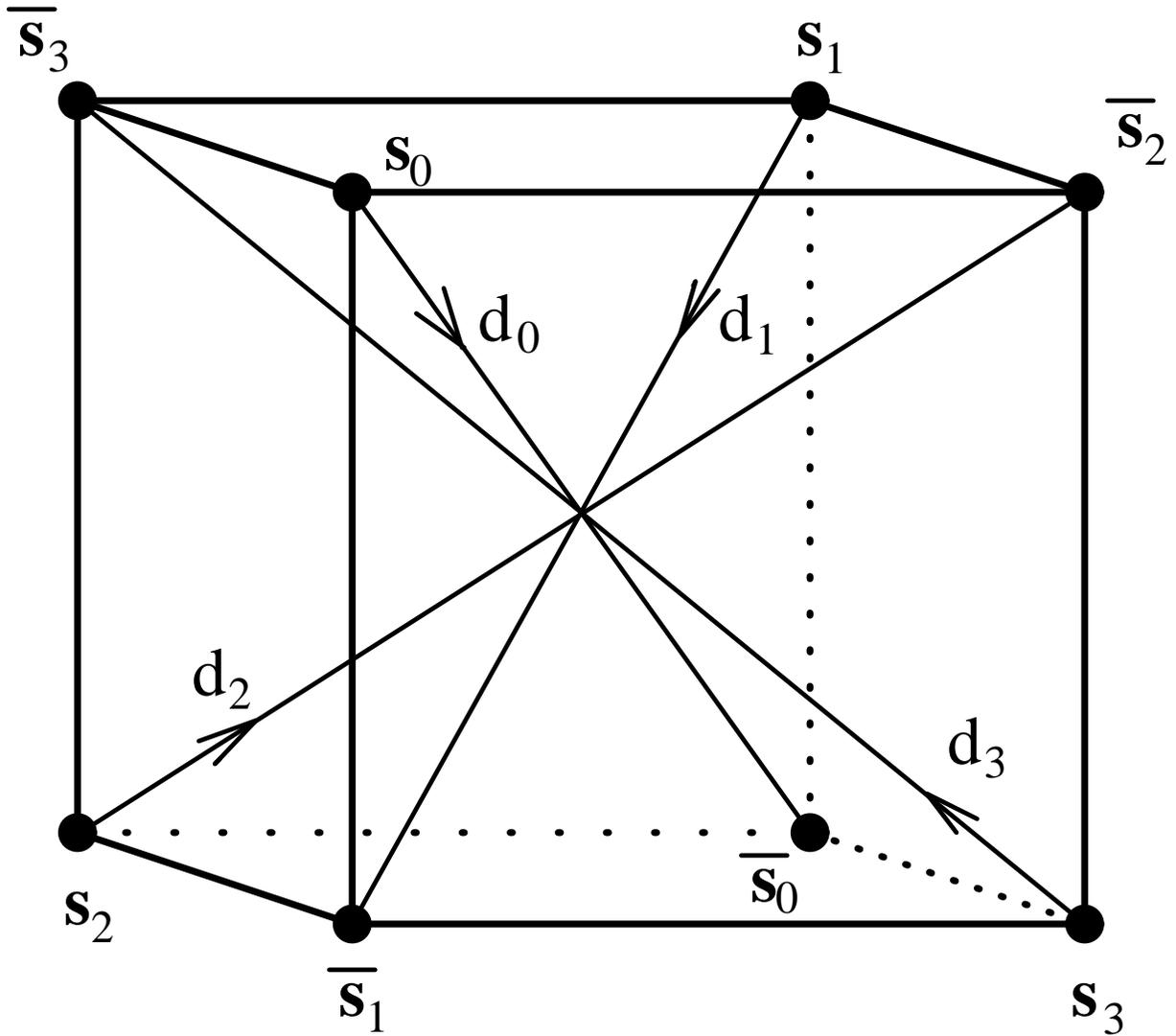
\bigskip Fig 2.  Arrangements of the spins $s_0,\ldots
,{\overline
s}_0$ on the corner sites of an elementary cube\break\indent of the simple
cubic
lattice $\cal L$;  $d_0,\,d_1,\,d_2,\,d_3$ denote four oriented diagonals
of the
cube.

\bigskip Fig.3. A typical elementary cube of ${\cal L}$, with corner spins
$\ss$ and the center spin~$\s$.

\end